# Self-aligned photonic defect microcavities with site-controlled quantum dots

*C.-W. Shih,* [1] *I. Limame,* [1] *C. C. Palekar,* [1] *A. Koulas-Simos,* [1]

*A. Kaganskiy,* [1] *P. Klenovský,*[2,3,b)] *S. Reitzenstein*[1,a)]

[1]*Institut für Festkörperphysik, Technische Universität Berlin, 10623 Berlin, Germany*

[2]*Department of Condensed Matter Physics, Faculty of Science, Masaryk University, Kotlářská 267/2, 61137 Brno, Czech Republic*

[3]*Czech Metrology Institute, Okružní 31, 63800 Brno, Czech Republic*

Correspondence:

a) S. Reitzenstein, E-Mail: stephan.reitzenstein@physik.tu-berlin.de
b) P. Klenovský, E-Mail: klenovsky@physics.muni.cz

**ABSTRACT:**

Despite the superiority in quantum properties, self-assembled semiconductor quantum dots face challenges in terms of scalable device integration because of their random growth positions, originating from the Stranski-Krastanov growth mode. Even with existing site-controlled growth techniques, for example, nanohole or buried stressor concepts, a further lithography and etching step with high spatial alignment requirements is necessary to accurately integrate QDs into the nanophotonic devices. Here, we report on the fabrication and characterization of strain-induced site-controlled microcavities where site-controlled quantum dots are positioned at the antinode of the optical mode field in a self-aligned manner without the need of any further nano-processing. We show that the Q-factor, mode volume, height, and the ellipticity of site-controlled microcavities can be tailored by the size of an integrated $AlAs/Al_2O_3$ buried stressor, with an opening ranging from 1 to 4 μm. Lasing signatures, including super-linear input-output response, linewidth narrowing near threshold, and gain competition above threshold, are observed for a 3.6-μm self-aligned cavity with a Q-factor of 18000. Furthermore, by waiving the rather complex lateral nano-structuring usually performed during the fabrication process of micropillar lasers and vertical-cavity surface emitting lasers, quasi-planar site-controlled cavities exhibit no detrimental effects of excitation power induced heating and thermal rollover. Our straightforward deterministic nanofabrication concept of high-quality quantum dot microcavities integrates seamlessly with the industrial-matured manufacturing process and the buried-stressor techniques, paving the way for exceptional scalability and straightforward manufacturing of high-β microlasers and bright quantum light sources.

**KEYWORDS** site-controlled quantum dots, photonic microcavity, nanolasers, microlasers, vertical-cavity surface-emitting lasers, scalable quantum light sources, buried-stressor method

## INTRODUCTION

Semiconductor quantum dots (QDs), acting as artificial atoms, have demonstrated superiority in a variety of emerging quantum technologies due to their exceptional optical properties. These include adjustable emission wavelength covering from the visible range to telecommunication bands[1], near-unity quantum efficiency[2], on-demand single-photon and entangled photon pair emission[3–5] for applications, for instance, in photonic quantum information technology[6]. Furthermore, self-assembled QDs enable the realization of cavity-enhanced low-threshold nanolasers with high-$\beta$ factors[7–9].

Nevertheless, standard QDs face significant challenges in epitaxially controlling their spectral inhomogeneous broadening and spatial nucleation position due to the self-assembled nature of Stranski-Krastonov (SK) growth. These issues hinder the scalable integration of QDs into nanophotonic structures, which aim to enhance the light extraction, emission rate, and light-matter coupling for optimum device performance. Spectral matching of single QDs and the optical modes of microresonators has been successfully addressed with multiple techniques including piezoelectric strain tuning[10–12], quantum confined Stark tuning[13,14], and temperature tuning[15]. The spatial matching, on the other hand, is addressed through post-growth deterministic nano-fabrication techniques like in-situ or marker-based optical and electron beam lithography (EBL)[16–18], pre-growth nanoholes patterning[19], and the buried-stressor growth method[20,21].

The buried-stressor growth technique for site-controlled quantum dots (SCQDs) is particularly intriguing due to its ability to ensure high optical quality of the QDs thanks to the high-quality growth surface[22,23], control of their position and local density[24,25], and seamless integration within the industry-scale vertical cavity surface emitting lasers (VCSELs) process flow[26]. Photonic devices, such as single-photon sources (SPS)[22–24,27] and micropillar lasers[25] based on SCQDs have been demonstrated. Nevertheless, the buried stressor method still requires a post-position-alignment step for the integration of SCQDs into photonic microcavities, which involves either marker-based or in-situ lithography[28]. The situation can become even more complex when the pre-patterned markers are overgrown and covered by thick layers such as distributed Bragg reflectors (DBRs), which could obstruct the visual identification of the markers[24,25].

QDs are often integrated into micro- and nano-cavities with small mode volume to harness the cavity enhancement originating from, for instance, the Purcell-effect in the weak coupling regime of cavity quantum electrodynamics. Despite the achieved performance, for instance, photon extraction efficiency exceeding 65% in micropillar[29,30] and 85% in circular Bragg grating (CBG) resonator configurations[5], accompanying issues related to thermal management[31,32] and detrimental surface defect states[33] arise. Microcavities based on buried photonic defects, where no deep etching is required, can overcome these issues, and have recently caught significant interest from the nanophotonics community[34–37]. In such quasi-planar microcavities, usually the upper DBRs is shaped into micro- or nano-lenses to laterally confine light.

In this study, we present an innovative type of quasi-planar site-controlled microcavity (SCM) created through the buried-stressor method. This SCM forms in a self-assembled manner during epitaxial growth and is spatially self-aligned to the SCQDs, eliminating the need for any post-growth lithography. We take advantage of the fact that the strain induced by the buried stressor can modulate the growth rate and reshape the morphology of the sequential layers including the upper DBR, and can therefore form a photonic defect providing strong lateral mode confinement. Interestingly, as we show below microcavity properties such as Q-factor and mode volume can be tailored by the geometry of SCM, which is controlled by the size of the oxide aperture.

In the following, we first describe the fabrication and formation of SCMs and SCQDs by means of the buried-stressor method, together with a theoretical investigation on the strain and optical aspects of SCMs. Afterward, we present the experimental characterization of SCMs with pump power dependent micro-photoluminescence measurements on SCMs with different full width at half maximum (FWHM). Finally, we demonstrate lasing action of SCM, proving the high potential of our approach for nanophotonic applications.

## RESULTS AND DISCUSSION

### Fabrication of SCMs

A crucially appealing aspect of the SCM concept is the fact that the related device fabrication is not only synchronized but also spatially self-aligned with the buried-stressor growth of SCQDs. The fabrication process can be divided into three major steps as illustrated in Fig. 1. It starts with (a) the 1$^{st}$-step metal-organic chemical vapor deposition (MOCVD) growth of oxidation templates with bottom DBR, followed by (b) mesa etching and lateral wet-oxidation of the AlAs layer to create the oxide aperture, and (c) the 2$^{nd}$-step overgrowth by InGaAs QDs and the top DBR.

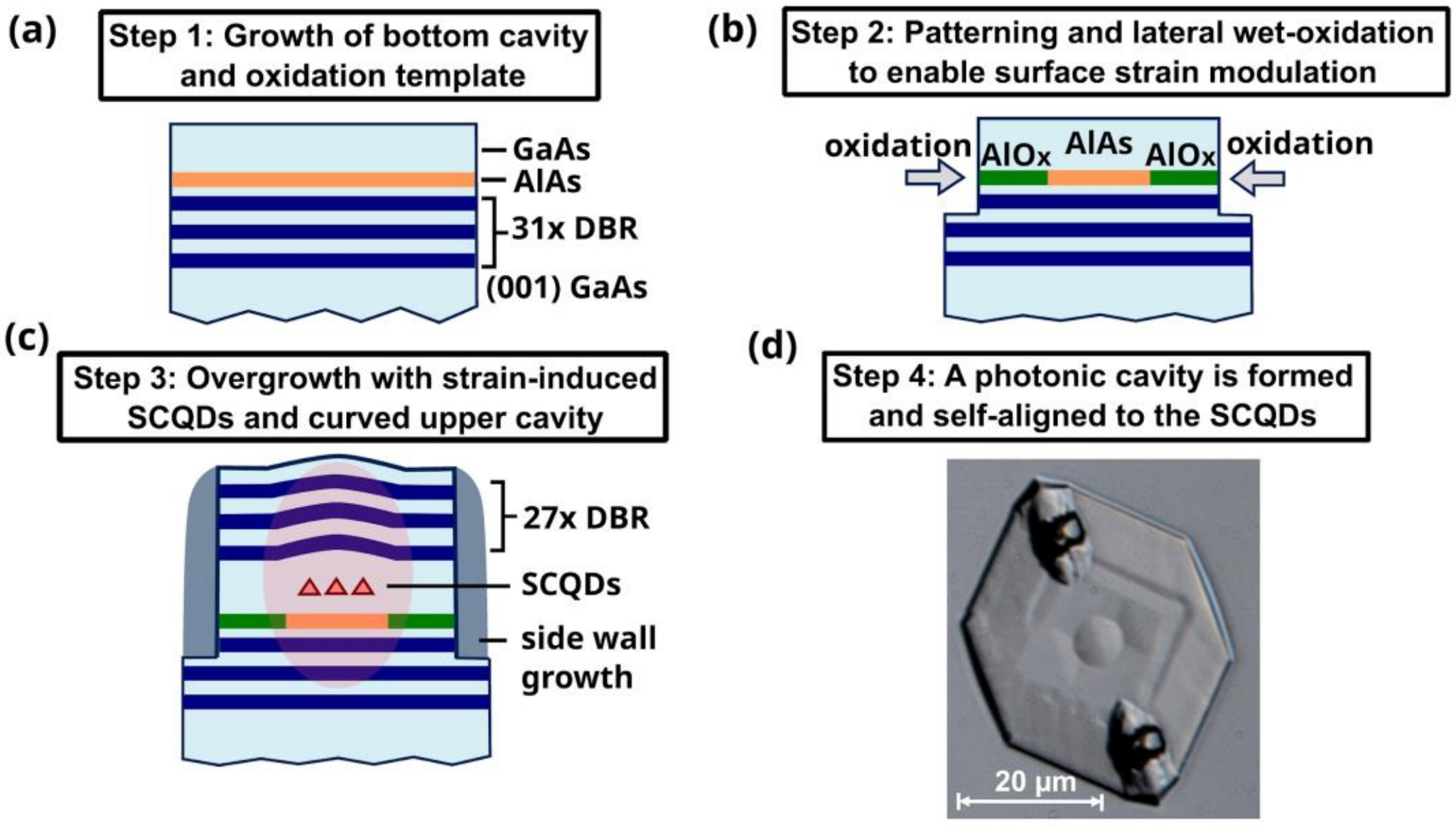

Fig. 1: Illustration of the fabrication process of SCMs with an overall flowchart on the left. (a) 1st-step epitaxial growth of oxidation templates. (b) Square mesa etching and partial lateral wet-oxidation to form a size- and site-controlled oxide aperture. (c) 2nd-step overgrowth with InGaAs QDs and top DBRs. The reddish ellipse depicts the mode confined by the photonic defect cavity. (d) Optical microscope image of a site-controlled microcavity self-aligned to the site-controlled QDs.

In our device structure discussed here, the bottom (top) DBR consists of 31 (27) alternating $Al_{0.9}Ga_{0.1}As$/GaAs layer pairs with quarter-wavelength thickness of 77.8/66.7 nm. The oxidation aperture in Fig. 1 consists of a 30-nm-thick AlAs oxidation layer embedded in two 50 nm thick $Al_{0.9}Ga_{0.1}As$ layers for mechanical stabilization. In the growth of the oxidation templates, 50 nm of GaAs is deposited on the surface to encapsulate the AlAs layer to prevent surface oxidation, as illustrated in Fig. 1 (a). Square mesas with a side length of approximately 20 μm are then patterned via UV-lithography using a Hg lamp before being exposed to water vapor at 420 ℃, laterally oxidizing the AlAs layer as illustrated in Fig. 1 (b). The lateral oxidation process is interrupted before fully oxidizing the AlAs layer, leaving an opening which we denote as oxide aperture. Systematic control of the oxide aperture is enabled by varying the side length of the square mesas in 67-nm steps in conjunction with in-situ optical microscope monitoring of the oxidation process. In the consecutive 2nd-step overgrowth, the InGaAs QDs are preferentially grown above the tensile-strained aperture because of the surface strain modulated by the oxide aperture[20,21]. Interestingly, in our SCM concept we utilize the induced strain not only for spatially controlling the QD nucleation site but also the thickness of all the subsequent layers, including the top DBR. During the deposition process of the upper DBR, the growth rate and thickness is modulated in a spatially dependent manner, causing the mirror layers of the top DBR above the oxide aperture to curve into a lens shape as illustrated in Fig. 1 (c) along with a microscope image in Fig. 1 (d).

Noteworthy, this industry-compatible process without any further post-growth lithography is sufficient to fabricate the SCM devices with 3D mode confinement.

**Morphology of SCM and theoretical investigations**

To verify the strain-modulated growth of the upper DBR, we performed atomic force microscopy (AFM) to scan the surface morphology of SCMs with various oxide aperture sizes as presented in Fig. 2 (a)-(c). Micro- and nano-lenses, isolated and surrounded by trenches, can be observed in the center of the mesas, i.e. directly above the oxide aperture. Please note that an ellipticity in the shape of SCM is observed, which can lift the degeneracy of the fundamental cavity mode and lead to a mode energy splitting as observed in bimodal micropillar lasers with asymmetric cross-section[38,39].

For a better understanding of the physics related to the results of the AFM measurements, we performed strain tensor calculations on the conceived structure depicted in Fig. 1 (c) using continuum elasticity theory [40–42]. Since the AFM measurements provide information about relative height in different parts of the surface, we focused on the calculations on vertical strain, i.e., $\varepsilon_{zz}$ strain component, where $z$ is oriented along the [001] crystal direction. The parameter $\varepsilon_{zz}$ then reveals the expected out-of-plane expansion or contraction of the crystal lattice towards the surface, which is then reflected by the surface morphology observed with AFM. The computed surface strain $\varepsilon_{zz}$ in panels (d)-(f) highly resembles the AFM profiles in (a)-(c), with an increase of the $\varepsilon_{zz}$ in the center, directly above the aperture, surrounded by diminishing $\varepsilon_{zz}$. Further details on our numerical calculations are provided in the Methods section.

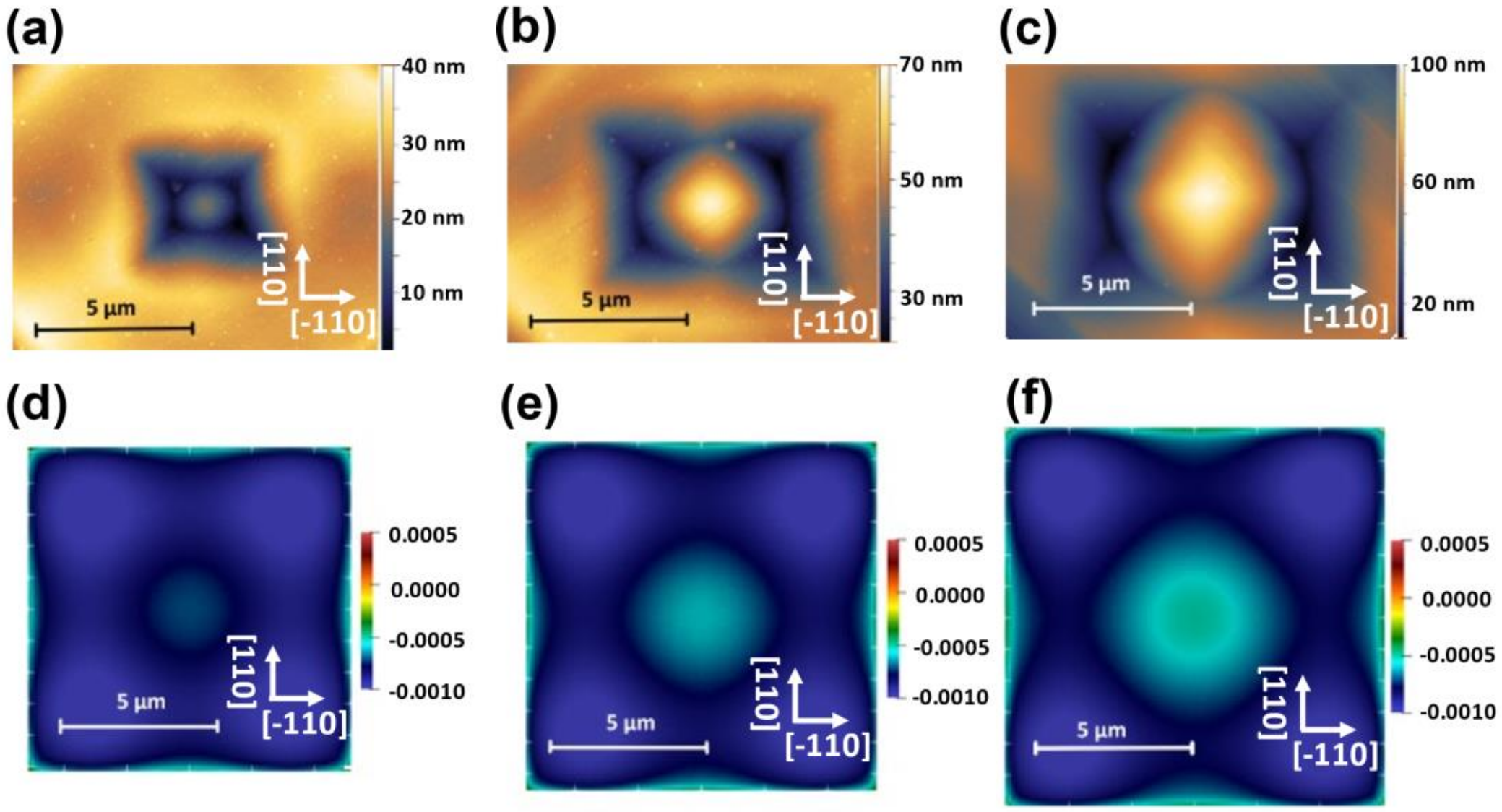

Fig. 2: AFM images of SCMs with FWHM of (a) 1.1 µm, (b) 2.4 µm, and (c) 3.6 µm. Note that the FWHM here is determined by fitting a Gaussian around the central bump to represent the size of the SCMs, which is explained later in Fig. 3. We also show the computed surface distributions of the *z* component of the strain tensor $\varepsilon_{zz}$ for aperture sizes of (d) 0.2 µm, (e) 1 µm, and (f) 2 µm. The range of magnitudes of $\varepsilon_{zz}$ strain in (d)-(f) is between -0.1% (violet) to -0.02% (green). Note the relative increase of $\varepsilon_{zz}$ above aperture (light blue) with respect to surrounding surface (violet) and correspondence to AFM images in (a)-(c).

We further analyzed the peak-to-valley height of the lenses as a function of their FWHM, fitted with a Gaussian lineshape, as exemplified in Fig. 3 (a). Interestingly, the extracted height increases linearly with the FWHM as shown in Fig. 3 (b), which can potentially affect the strength of lateral confinement.  Due to the lack of circular symmetry of the SCMs (see AFM images in Fig. 2), their height and FWHM are estimated by averaging two orthogonal line profiles per AFM image. Meanwhile, the AFM surface characterization of the SCQDs layer underneath the top DBRs has revealed an additional abrupt small bump around 8 nm existing above the oxide aperture, which can be attributed to Indium migration during epitaxial growth[43] (please see the Supplementary Information). Similar behavior is also observed in the theoretical surface strain calculations as shown in Fig. 3 (c). Here, we compute the strain increase $\Delta\varepsilon_{zz}$ at the center of the sample surface {see Fig. 2 (d)-(f)} with respect to the surrounding area with smallest $\varepsilon_{zz}$ magnitude in analogy to Fig. 3 (a) and plot it as a function of the corresponding $\varepsilon_{zz}$ FWHM. The high correspondence between the AFM profiles and the calculated demonstrates that the whole crystal lattice of the upper layers is distorted due to the buried stressor. Overall, the lens-shape morphology indicates that in our self-aligned fabrication concept not only the growth of the SCQDs, but also the cavity geometry can be tailored and controlled by the size and shape of the oxide aperture.

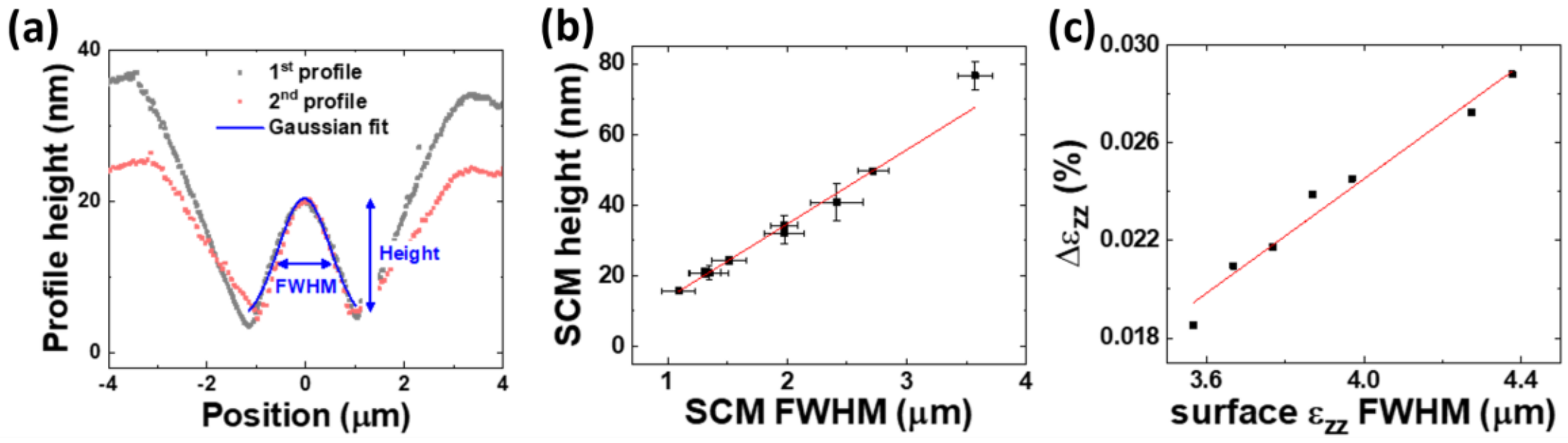

Fig. 3: (a) Exemplary Gaussian fit of the AFM profile of the SCM presented in Fig. 2 (a). (b) The peak-to-valley height of SCM as a function of the FWHM. The presented SCM FWHM and height are averaged from two orthogonal line profiles. (c) The same as in (b) but derived from the $\varepsilon_{zz}$ strain computed by the continuum elasticity theory.

Figure 4 (a) presents a cross-section scanning-electron microscopy (SEM) image of a structure with a notably large oxide aperture. A kink, traced by an arrow, along the upper layers of the SEM image implies how the strain exerted by the $AlAs/Al_2O_3$ aperture propagates towards the surface. Due to the oxidation inhomogeneity and the lack of direct measurement of the absolute oxide aperture size after overgrowth, we deduce the oxide aperture diameter from the measured SCM FWHM as presented in Fig. 4 (b), assuming that the estimated propagation angle of 65° is independent of the oxide aperture size. This estimation is supported by the variation of the SCM

FWHM, which is equal to the nominal variation of the square mesa size in another more homogenous array. That array, however, has a smaller range of oxide aperture size variation, purely limited by the design of our UV-lithography mask as shown in Fig. 4 (c). Although there are a few outliers resulting from oxidation inhomogeneity, it is evident that the SCM FWHM can be controlled by the oxide aperture size, i.e., the mesa size.

To gain more insights, we performed theory calculations of $\varepsilon_{zz}$ strain, as presented in Fig. 4 (d). An excerpt of the simulated structure in (d) corresponds in all dimensions to that for experimental SEM in (a). The calculation clearly indicates that the magnitude of $\varepsilon_{zz}$ in GaAs DBR layers increases from more (white) towards less (dark grey) compressive strain as one moves from the AlAs/$Al_2O_3$ oxide aperture towards the sample surface. Moreover, the magnitude of $\varepsilon_{zz}$ strain increases in "cone" similarly as in our SEM image. However, the "kink" on the sample surface layer is related to a more gradual change of $\varepsilon_{zz}$ as indicated by the inset arrow. Furthermore, in Fig. 4 (e) we show the aperture diameter as a function of $\varepsilon_{zz}$ FWHM obtained from our calculations. We find that the experimental and theoretical observations in (b) and (e) are in good qualitative agreement, and that the aperture diameter increases linearly with slopes of $\approx$1.7 and $\approx$2.0, respectively. However, there is a clear vertical offset between the curves by about 2 μm, which can be attributed to multiple effects: firstly, the surface AFM profiles account for the accumulation of the thickness changes in all the layers with each corresponding strain instead of only the surface strain as presented in (e); secondly, the experimental growth rate of GaAs and $Al_{0.9}Ga_{0.1}As$ can also be modulated by the surface strain, leading to a more complex picture; thirdly, the lateral expansion of strain exerted by the aperture towards the sample surface would also distort the lateral dimension.

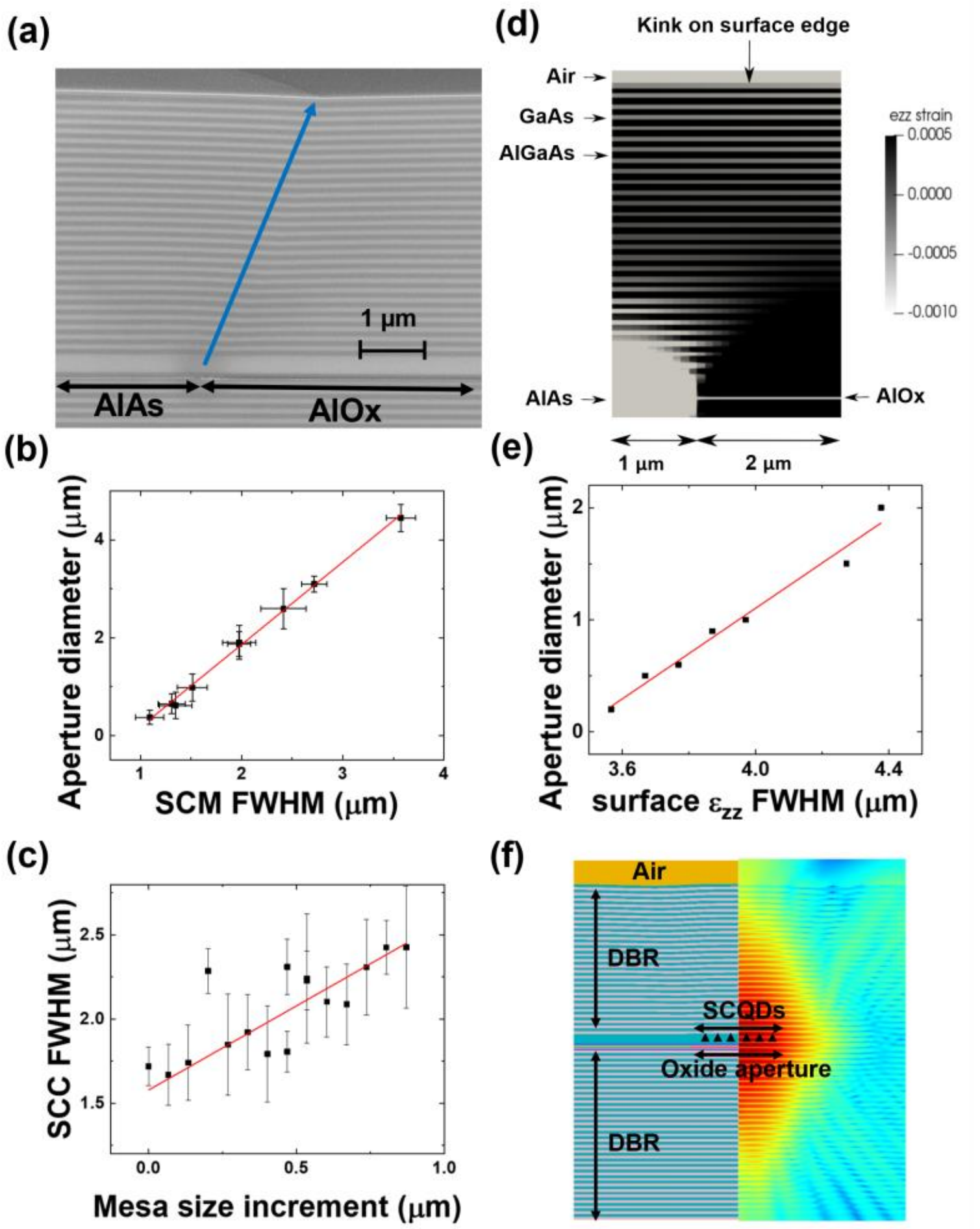

Fig. 4: (a) Cross-section SEM image of an SCM showing the propagation of the profile kink from the oxide aperture edge. (b) Estimated oxide aperture diameter as a function of SCM FWHM. (c) SCM FWHM as a function of the nominal mesa size increment defined by UV-lithography. The red linear line is a plot with a slope of 1 and y-intersection at 1.58 µm as a guide to the eye. (d) Cutout from the strain calculation corresponding to the SEM measurement (a). The gray DBR layers in (d) correspond to GaAs while the black ones to $Al_{0.9}Ga_{0.1}As$. The kink on the surface seen in (a) occurs as a gradual lateral change of $\varepsilon_{zz}$ magnitude on the top sample surface layer and is indicated by an inset arrow in (d). (e) The same as in (b) but derived from $\varepsilon_{zz}$ calculations. (f) Numerical simulation of the optical properties of an SCM with an aperture diameter of 2.5 µm. The opening of the oxide aperture is denoted as the SCQDs growth area, self-aligned to the antinode of the SCM mode field.

To investigate the optical properties of SCMs, optical simulations are carried out by means of a commercial software package, JCMsuite, based on the finite element method (FEM)[44]. The modeling of the structure is described in the Methods section. Figure 4 (f) shows the optical field of the fundamental cavity mode confined by an SCM with an exemplary aperture diameter of 2.5 μm. The mode field extends out of the oxide aperture, resulting in a strong spatial overlap between the SCQDs and the mode field antinode, which maximizes the light confinement and the Purcell factors of the integrated QD emitters. Furthermore, the self-aligned overlap of SCQDs and modes is advantageous for improving the interaction between light and matter and for minimizing the absorption losses attributed to off-center QDs[25]. The simulation yields a theoretical cavity Q-factor of 144 000 and mode volume of 44.1 $(\lambda/n)^3$, which would result in a Purcell factor of 248 in the ideal case of perfect spatial and spectral alignment. We would, however, like to remark that the mode parameters of such buried photonic cavities are highly sensitive to the shape and morphology of the defect structure. For instance, a Gaussian-shape defect cavity would have a significantly higher Q-factor than a simple mesa defect cavity[35]. Therefore, without using nano-resolution imaging techniques such as the transmission electron microscopy (TEM) to accurately retrieve the layer structures for modeling, the presented simulation is expected to deviate from the experiments considerably.

**Optical characterization of SCMs**

We experimentally characterized the optical properties of SCMs using a micro-photoluminescence (μPL) setup, where the sample is mounted in a He-flow cryostat at 20 K and optically pumped by a continuous wave (CW) laser at 785 nm followed by a spectrometer with a spectral resolution of 55 μeV. Figure 5 shows pump-power and polarization dependent measurements of an SCM with a FWHM of 2.7 μm. Interestingly, two ensembles of optical modes can be observed in the μPL spectra, as shown in Fig. 5 (a) - (c). We attribute the presence of these two groups of modes to lateral light confinement arising from the strain-induced aperture cavity and the mesa cavity, respectively. Here, tight lateral light confinement in the strain-induced aperture cavity is caused by the curved layers of the top DBR directly above the oxide aperture, which we denote as SCM throughout the article. The overgrown etched square mesa with a length of 20 μm also leads to lateral mode confinement, similar to other previous etch-then-deposit approaches[36,37]. However, the associated mode volume is significantly larger, and as a result, compared to the aperture cavity the mesa cavity leads to smaller mode spacing as a sign of larger mode volume. Observed typical mode spacings of the aperture cavity and the mesa cavity are 1-10 meV and 0.2 meV, respectively. Noteworthy, the aperture cavity modes are spatially centered and self-aligned to the SCQDs. On the other hand, because of deviations of the structural geometry, the mesa cavity modes are typically located slightly off-center of the oxide aperture by more than 2 μm and are therefore spatially mismatched to the SCQDs.

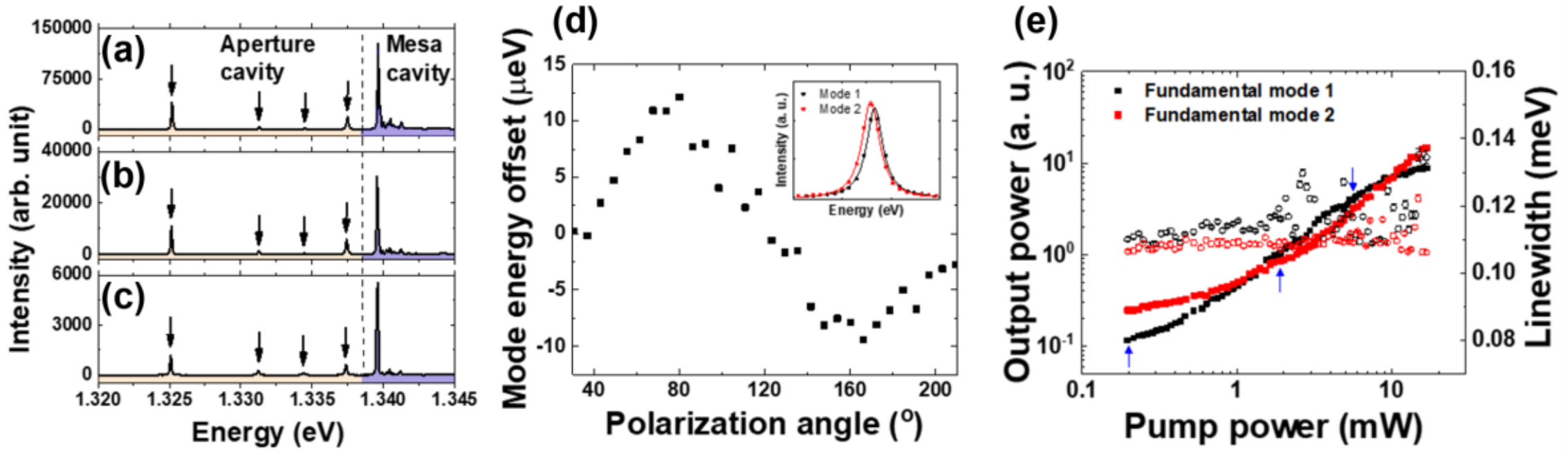

Fig. 5: Pump power and polarization dependent μPL measurements of an SCM with a FWHM of 2.7 μm. (a)-(c) μPL spectra of fundamental mode 1 collected at the center of the mesa at pump powers of 5.6 mW, 1.9 mW, and 0.2 mW, respectively. The arrows indicate the aperture cavity modes. (d) The aperture mode energy fitted as a function of the polarization angle. The inset shows the spectra of the two orthogonal polarized fundamental modes. (e) Pump power dependent output intensity and linewidth of the aperture

cavity mode. The solid squares belong to the output power and the open circles belong to the linewidth. The arrows indicate the powers corresponding to (a)-(c).

We would like to note that the strain-induced aperture cavity modes have a lower energy than the mesa cavity modes. That, on the first sight, surprising feature, can be attributed to the fact that the aperture cavity modes are confined in vertical direction by a larger thickness of the central cavity region above the oxide aperture. Here, the larger thickness is confirmed by our AFM measurements (see Fig. 2 and the Supplementary Information). Since the vertical mode confinement is stronger than the lateral one, small changes in the cavity thickness have a strong influence on the mode energy. The mentioned mode confinement effects are further verified by 1-D numerical simulations with the transfer matrix method (TMM), investigating the mode energy on and off the oxidized area. The simulations show that a simple GaAs cavity just 8 nm thicker with a non-oxidized AlAs layer can redshift the mode energy by about 10 meV, fitting our experimental observation. Please see the Supplementary Information for more details on the TMM simulation. Furthermore, with polarization dependent measurements we find that the non-circular shape of the aperture cavity also lifts the degeneracy of the fundamental mode, leading to two orthogonally polarized modes with a spectral splitting of 20 μeV as shown in Fig. 5 (d). Such a phenomenon has also been observed in bimodal micropillar lasers with asymmetric cross-section[38,39]. These two mode components are marked as fundamental mode 1 and mode 2 in Fig. 5 (e). Note that the mesa cavity modes also exhibit a high degree of linear polarization (please see the Supplementary Information). Without intentionally increasing the number of SCQDs, for instance, by stacking multiple layers of high density SCQDs[43], most of the oxide apertures contain a low number of SCQDs. In fact, less than 20 SCQDs are formed for mesas with oxide aperture diameter in the range of 700 to 1400 nm in our sample[25]. As a result, most of the structures do not exhibit nonlinear lasing signatures or power dependent linewidth narrowing due to the inadequate modal gain to surpass the laser threshold. These excitation power dependent properties are exemplified in Fig. 5 (e).

To obtain a better understanding of the SCM's optical properties, we systematically investigated SCMs with different FWHM. Figure 6 (a) shows the cavity Q-factor as a function of SCM FWHM evaluated at low pump power. The Q-factor increases from around 8000 to 18000 when the FWHM increases from 2.0 µm to 3.6 µm. Below 2 µm no clear trend in Q-factor is observed, potentially due to the complexity of the overall SCM shape accompanying the FWHM variation as shown in Fig. 2. The mode energy, on the other hand, strongly blueshifts with decreasing SCM FWHM as shown in Fig. 6 (b). That effect can be attributed to two factors: firstly, the mode radius decreases with decreasing SCM FWHM[45]; and secondly, the optical cavity thickness is also reduced as presented in Fig. 2. Figure 6 (c) shows the 1$^{st}$-2$^{nd}$ mode and fundamental mode splitting as a function of SCM FWHM. The increase of 1$^{st}$-2$^{nd}$ mode splitting with decreasing SCM FWHM is a further experimental indicator that the mode is laterally confined and controlled by the SCM FWHM, and therefore by the size of the oxide aperture, with reference to the case of cylindrical Fabry-Pérot microcavities[45]. Note that below 2 µm FWHM, along with the large amount of blueshift and a 1$^{st}$-2$^{nd}$ mode splitting exceeding 9 meV, the higher order modes of the SCMs are spectrally indistinguishable from the mesa cavity mode ensemble. Moreover, the splitting between the two orthogonally polarized fundamental modes is also affected by the SCM's FWHM. As the SCM becomes smaller, the non-circular shape leads to a significant splitting of up to 163 µeV. The amount of splitting reduces with increasing SCM size, indicating the reduction of SCM ellipticity.

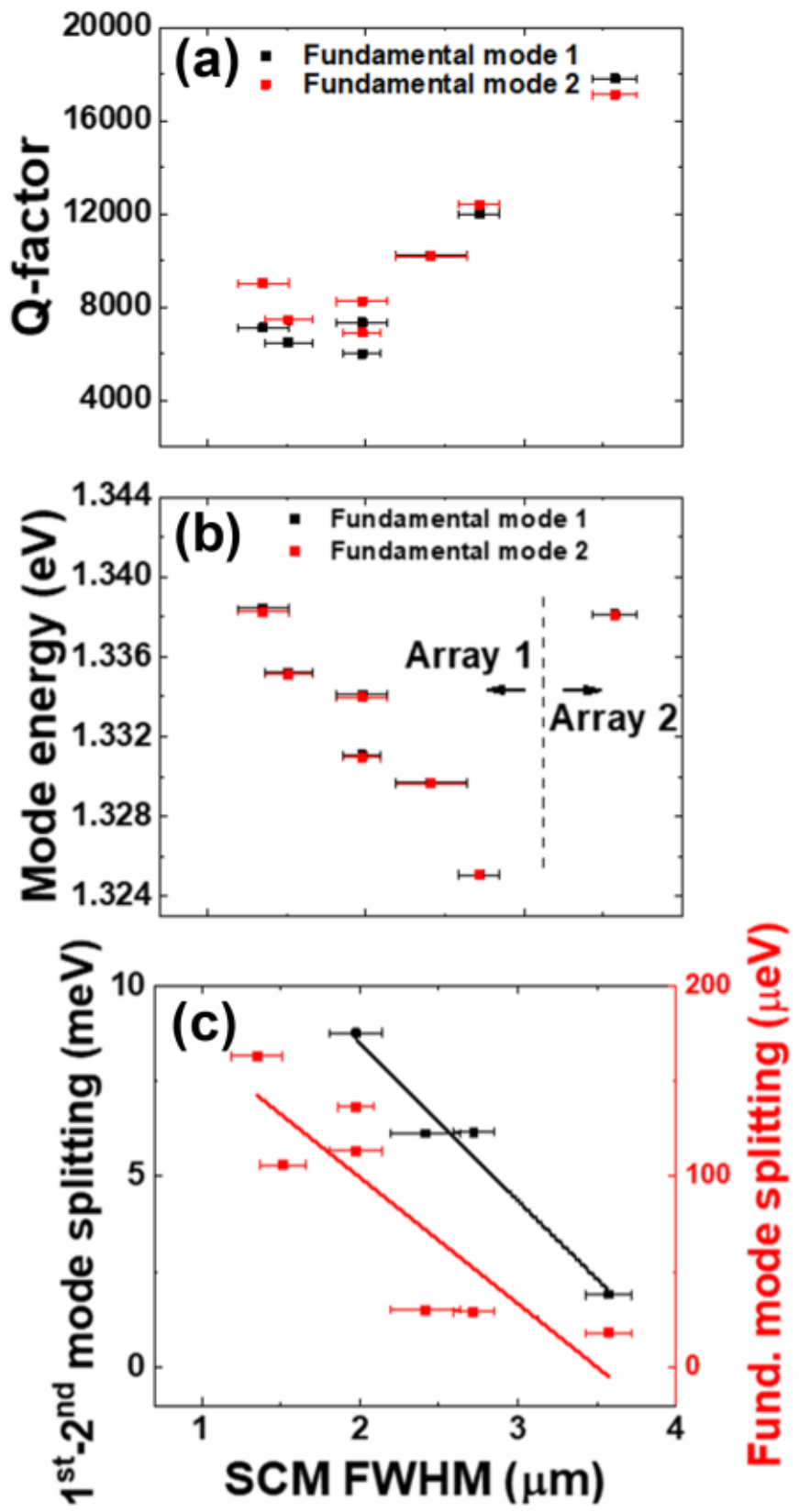

Fig. 6: Cavity (a) Q-factor, (b) mode energy, and (c) mode splitting as a function of the FWHM. Note that the SCM with a FWHM of 3.6 μm locates at another array than the others on the wafer, where the epitaxial inhomogeneity of the layer thickness dominates over the local effects and therefore shifts its mode energy out of the expected trend.

The developed high-quality SCMs have high potential to act as QD microlasers with nanoengineered gain medium. To study lasing action of our SCMs, we selected a structure with 3.6 μm FWHM where we expected many SCQDs. We estimate the number of SCQDs to surpass 60, extrapolated from the previous study[25], interacting with the optical field of the cavity to provide sufficient optical gain to overcome the losses. The excitation-power-dependent spectra are presented in Fig. 7 (a), where single mode lasing with the fundamental mode is observed. The SCM shows lasing signatures, including a characteristic S-shaped IO-curve in double-logarithmic scale and linewidth narrowing near threshold reaching the resolution limit of 55 μeV of our spectrometer as depicted in Fig. 7 (b). Similar to in the case of bimodal micropillar lasers, gain competition is also presented between the two orthogonal components of the fundamental modes, where mode 1 is the superior component. By means of a simple laser rate equation assuming a simple two-level system[46], we fitted the IO-curve of the superior lasing mode to extract a lasing threshold $P_{th}$ of (6.7 ± 0.3) mW, defined as the threshold where the mean photon number reaches unity, and a spontaneous emission factor β of (4.2 ± 0.1) × $10^{-4}$.

Another interesting feature of the SCM lasing performance is the excitation power dependence of the mode energy. Fitting the experimental emission spectra yields a pump power dependent blueshift of around 260 μeV in the range from 0.5 mW to 30 mW, as shown in Fig. 7 (c). This spectral shift can be attributed to the carrier-induced refractive index change, known as plasma effects[47,48]. We emphasize that the DBRs in the structure consist of $Al_{0.9}Ga_{0.1}As$ and GaAs, which can optically absorb the pump laser at 785 nm, leading to low pump power conversion efficiency[32]. In the case of deeply etched structures, such as micropillar lasers, this usually leads to heating of the devices, which induces undesired thermal effects such as a redshift of the mode energy[31,32]. Noteworthy, such heating effects are not observed in our quasi-planar SCM, where the heat can be effectively dissipated with the non-isolated surroundings.

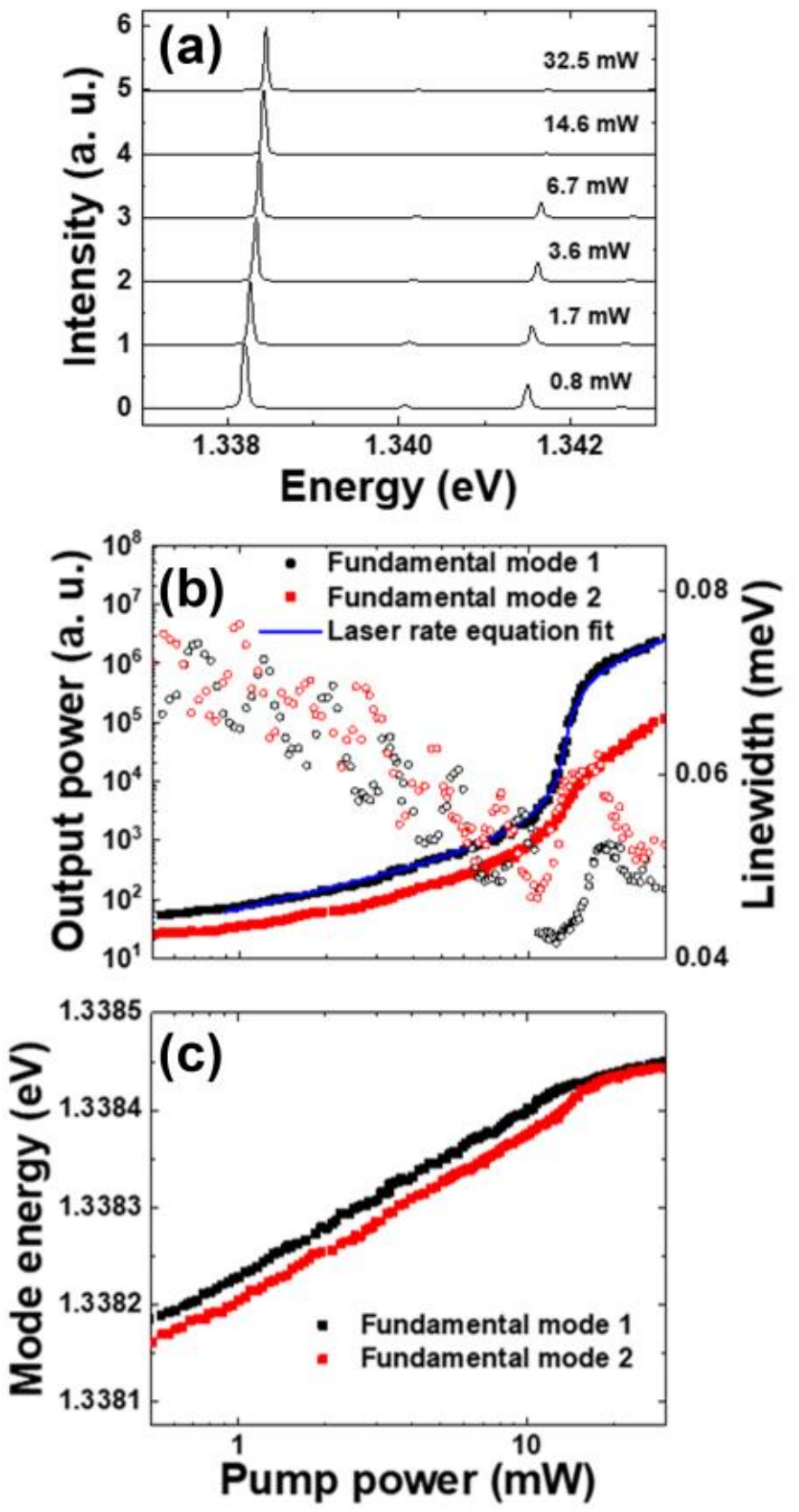

Fig. 7: Pump power dependent measurement of a SCM with 3.6 μm FWHM. (a) The pump power dependent spectra with normalized intensity. (b) IO-curve and linewidth of both components of the fundamental modes. The solid markers represent the output powers, whereas the empty markers denote the linewidths. (c) The pump power dependent mode energy.

Intriguingly, all the presented cavity Q-factors of SCMs are comparable or even higher than for the case of free-standing micropillars with a diameter of 5.2 μm and slightly larger mode volume on the same wafer, where the achieved cavity Q-factor is lower than 10000 [25]. The lasing threshold (6.7 ± 0.3) mW of the SCM microlaser is similar to the values determined for 5.2-μm-diameter micropillar lasers, where the achieved lasing threshold ranges from 5 to 9 mW [25]. This proves the high quality of our SCMs and their potential for high-performance quantum light sources and micro-lasers applications. We would like to note that not only the SCM mode energy but also the SCQDs emission energy are strongly influenced by the oxide aperture[43], which can lead to a spectral mismatch between the two. Together with the strongly absorbing GaAs/$Al_{0.9}Ga_{0.1}As$ DBRs and the low number of SCQDs in most of the structures, the lasing action is limited. However, with further efforts in optimizing all the aforementioned aspects, including stacked layers of SCQDs and replacing the DBRs with low-absorbing $Al_{0.2}Ga_{0.8}As/Al_{0.9}Ga_{0.1}As$ ones[32,43], we expect the SCM lasers' performance in terms of $P_{th}$ and $\beta$ can be hugely improved.

## CONCLUSIONS

In conclusion, we showed that by means of the buried-stressor growth method, not only SCQDs are preferably nucleating above the oxide aperture, but also SCMs are formed self-aligned to the SCQDs resulting from the strain-induced growth-rate modulation in the upper DBR. The latter leads to a tight 3-D mode confinement with high cavity Q-factor and small mode volume without further post-growth lithography. Our experimental characterization reveals that the optical properties of SCMs such as Q-factor, mode energy, and mode volume are strongly influenced by the size of the oxide aperture. Lasing, including a characteristic S-shaped I/O-curve in double-logarithmic scale and linewidth narrowing, was achieved for an SCM with a Q-factor of 18000 and a rather large FWHM of 3.6 μm, which has enough modal gain provided by SCQDs to overcome the optical losses. Additionally, no thermal-induced redshift of the mode energy is observed for the whole pump power range, highlighting the effectiveness of thermal dissipation in such quasi-planar structures.

Thanks to the fact that the formation of SCMs is induced by the buried stressor, no nanoscale lithography technique such as EBL is required in the fabrication process. In fact, UV-lithography as a highly scalable and cost-effective lithography technique is sufficient to create these microcavities with 3D mode confinement. The SCM concept is well aligned with the established manufacturing processes of VCSELs and SCQDs, enabling an immediately applicable approach for the production of scalable and high-quality nanophotonic devices utilizing QDs.

## MATERIALS AND METHODS

### Structural modeling of SCM for optical simulation

The difficulties of theoretically investigating the SCMs' optical performances lie in the precise structural modeling of the SCM topology. It has been shown that for such photonic defect cavities, the cavity Q-factor and the mode volume are highly sensitive to the shape and size of the defects buried underneath the top DBRs[35]. In this work, we model the structures in the following way to match the available AFM and SEM profile: (a) We assume rotational symmetry in the simulation in order to have as simple a model as possible, but of course never simpler. (b) A small Gaussian bump with a height of 8 nm and a standard deviation of oxide aperture radius is introduced in the GaAs cavity layer before the top DBR as evidenced by the AFM profile on the SCQDs layer. Please see the Supplementary Information for the AFM profile on the SCQDs layer. (c) We modify the thickness of each top DBR according to an inverse Gaussian function: $t(r) = T\times(1-A\times G(r, r_0))$, where $t(r)$ is the modulated thickness as a function of the radial position, $T$ is the nominal thickness of each layer, $A=0.12$ modifies the strength to match the peak-to-valley height from the surface AFM profile, and $G$ is a normalized Gaussian function centered at $r_0$ with a standard deviation of $r_0/4$. For each layer, we let $r_0$ to propagate at 65° angle from the edge of the oxide aperture according to the SEM image.

**Numerical modeling of elastic strain**

In this work, we used a continuum elasticity approach for modeling of the elastic strain in the structure using software Nextnano++[40]. The method starts with the definition of the simulated structure, which was in our case taken to be exactly the same as in Fig. 1 (c) including all dimensions therein. Each material in the structure is thereafter described by the corresponding elastic parameters from Nextnano++ software database. In case of alloys, a linear interpolation between parameters for constituent materials was employed. The elastic strain energy in the whole simulated structure was thereafter minimized on the finite difference grid. We note that the finite difference method was used in favor of potentially more precise atomistic approaches[41,49] because of the considerably large dimensions of the simulated structure, being on the order of several μm in each spatial extension.

**Optical measurements**

We investigate the structures with a standard μPL setup. The sample is mounted in a He-flow cryostat operating at 20 K, optically excited by a CW laser at 785 nm with a microscope objective of 0.4 numerical aperture to focus the laser beam into a small spot around 4 $\mu m^2$ and simultaneously collect the sample emission. In the detection path, a rotatable lambda-half wave plate followed by a fixed linear polarizer are installed before the spectrometer to resolve the polarization dependency.

**Laser rate equations model based on a two-level system**

To qualitatively characterize the lasing threshold $P_{th}$ and the β-factor of the SCM laser, a simple laser rate equation model based on a two-level system is applied[46]. By solving the laser rate equations, the optical pump power $P_{pump}$ can be expressed as a function of the output power $P_{out}$:

$$P_{pump}(A,B,\beta,\xi) = A\frac{\gamma}{\beta}\left[\frac{BP_{out}}{1+BP_{out}}(1+\xi)(1+BP_{out}) - \beta\xi BP_{out}\right],$$

where A is the pump scaling parameter, B is the output scaling parameter linking the intracavity photon number to the measured output power, $\xi = n_0\beta/\gamma\tau_{sp}$ is a dimensionless factor with $n_0$ the exciton number at transparency threshold, $\tau_{sp}$ the spontaneous emission rate, $\gamma$ the cavity decay rate, and $\beta$ the spontaneous emission factor. On the other hand, $P_{th}$ defined as the pump power when the mean photon number reaches unity and is expressed by:

$$P_{th} = A(\xi(1-\beta)+1+\beta)\gamma/2\beta$$

In the fit, we additionally take $\tau_{sp}$ = 1 ns and $n_0$ = 2900 as have been previously reported for QD micropillar lasers[50] to avoid the degeneracy of the fitting parameters.

## Acknowledgments

This work received funding from the German Research Foundation (Re2974/20-1, Re2974/21-1) and the Volkswagen Foundation (NeuroQNet II) and from the European Research Council under the European Union's Seventh Framework ERC Grant Agreement No. 615613. Furthermore, we acknowledge the projects SEQUME (20FUN05) and QADeT (20IND05) from the EMPIR program cofinanced by the Participating States and from the European Union's Horizon 2020 research and innovation program. P. K. was partly funded by the Institutional Subsidy for the Long-Term Conceptual Development of a Research Organization granted to the Czech Metrology Institute by the Ministry of Industry and Trade of the Czech Republic and by the project Quantum Materials for applications in sustainable technologies, CZ.02.01.01/00/22_008/0004572. In addition, we thank K. Schatke, R. Schmidt, and R. Linke for the technical support on epitaxy and cleanroom fabrication. We also acknowledge the members of our research groups for fostering a supportive and collaborative environment, especially, K. Gaur, J. Große, and T. Müller.

## Conflict of Interest

The authors declare no conflict of interest.

## Data availability

Data underlying the results presented in this paper may be obtained from the authors upon reasonable request.

## Author contributions

C.-W. S. conducted the measurements, analysis, and optical simulations. C.-W. S. and S. R. wrote the manuscript. I. L. provided preliminary results and concept. C. C. P. and A. K.-S. assisted the measurements. A. K. fabricated the sample. P. K. performed the strain simulation, included the corresponding theory, and revised the manuscript. S. R. acquired funding and supervised the project.

## Supplementary information

Supplementary information accompanies the manuscript on the Light: Science & Applications website (http://www.nature.com/lsa).

## References

1. Sittig, R. *et al.* Thin-film InGaAs metamorphic buffer for telecom C-band InAs quantum dots and optical resonators on GaAs platform. *Nanophotonics* **11**, 1109–1116 (2022).
2. Große, J., Mrowiński, P., Srocka, N. & Reitzenstein, S. Quantum efficiency and oscillator strength of InGaAs quantum dots for single-photon sources emitting in the telecommunication O-band. *Appl. Phys. Lett.* **119**, 061103 (2021).

3. Kolatschek, S. *et al.* Bright Purcell Enhanced Single-Photon Source in the Telecom O-Band Based on a Quantum Dot in a Circular Bragg Grating. *Nano Lett.* **21**, 7740–7745 (2021).

4. Shih, C.-W., Rodt, S. & Reitzenstein, S. Universal design method for bright quantum light sources based on circular Bragg grating cavities. *Opt. Express* **31**, 35552 (2023).

5. Liu, J. *et al.* A solid-state source of strongly entangled photon pairs with high brightness and indistinguishability. *Nat. Nanotechnol.* **14**, 586–593 (2019).

6. Heindel, T., Kim, J.-H., Gregersen, N., Rastelli, A. & Reitzenstein, S. Quantum dots for photonic quantum information technology. *Adv. Opt. and Photonics* **15**, 613 (2023).

7. Ota, Y., Kakuda, M., Watanabe, K., Iwamoto, S. & Arakawa, Y. Thresholdless quantum dot nanolaser. *Opt. Express* **25**, 19981 (2017).

8. Deng, H., Lippi, G. L., Mørk, J., Wiersig, J. & Reitzenstein, S. Physics and Applications of High-β Micro- and Nanolasers. *Adv. Opt. Mater* **9**, 2100415 (2021).

9. Kreinberg, S. *et al.* Emission from quantum-dot high-ß microcavities: Transition from spontaneous emission to lasing and the effects of superradiant emitter coupling. *Light Sci. Appl.* **6**, e17030 (2017).

10. Zhang, J. *et al.* High yield and ultrafast sources of electrically triggered entangled-photon pairs based on strain-tunable quantum dots. *Nat. Commun.* **6**, 10067 (2015).

11. Ding, F. *et al.* Tuning the exciton binding energies in single self-assembled InGaAs/GaAs quantum dots by piezoelectric-induced biaxial stress. *Phys. Rev. Lett.* **104**, 067405 (2010).

12. Aberl, J. *et al.* Inversion of the exciton built-in dipole moment in In(Ga)As quantum dots via nonlinear piezoelectric effect. *Phys. Rev. B* **96**, 045414 (2017).

13. Bennett, A. J. *et al.* Giant Stark effect in the emission of single semiconductor quantum dots. *Appl. Phys. Lett.* **97**, 031104 (2010).

14. Huang, H. *et al.* Electric field induced tuning of electronic correlation in weakly confining quantum dots. *Phys. Rev. B* **104**, 165401 (2021).

15. Reithmaier, J. P. *et al.* Strong coupling in a single quantum dot–semiconductor microcavity system. *Nature* **432**, 197–200 (2004).

16. Sapienza, L., Davanço, M., Badolato, A. & Srinivasan, K. Nanoscale optical positioning of single quantum dots for bright and pure single-photon emission. *Nat. Commun.* **6**, 7833 (2015).

17. Rodt, S. & Reitzenstein, S. High-performance deterministic in situ electron-beam lithography enabled by cathodoluminescence spectroscopy. *Nano Express* **2**, 014007 (2021).

18. Dousse, A. *et al.* Controlled light-matter coupling for a single quantum dot embedded in a pillar microcavity using far-field optical lithography. *Phys. Rev. Lett.* **101**, 267404 (2008).

19. Nakamura, Y. *et al.* Vertical alignment of laterally ordered InAs and InGaAs quantum dot arrays on patterned (0 0 1) GaAs substrates. *J. Cryst. Growth* **242**, 339–344 (2002).

20. Strittmatter, A. *et al.* Lateral positioning of InGaAs quantum dots using a buried stressor. *Appl. Phys. Lett.* **100**, 093111 (2012).

21. Strittmatter, A. *et al.* Site-controlled quantum dot growth on buried oxide stressor layers. *Phys. Status Solidi A* **209**, 2411–2420 (2012).

22. Strauß, M. *et al.* Resonance fluorescence of a site-controlled quantum dot realized by the buried-stressor growth technique. *Appl. Phys. Lett.* **110**, 111101 (2017).

23. Große, J., von Helversen, M., Koulas-Simos, A., Hermann, M. & Reitzenstein, S. Development of site-controlled quantum dot arrays acting as scalable sources of indistinguishable photons. *APL Photonics* **5**, 096107 (2020).

24. Kaganskiy, A. *et al.* Micropillars with a controlled number of site-controlled quantum dots. *Appl. Phys. Lett.* **112**, 071101 (2018).

25. Kaganskiy, A., Kreinberg, S., Porte, X. & Reitzenstein, S. Micropillar lasers with site-controlled quantum dots as active medium. *Optica* **6**, 404 (2019).

26. Huffaker, D. L., Deppe, D. G., Kumar, K. & Rogers, T. J. Native-oxide defined ring contact for low threshold vertical-cavity lasers. *Appl. Phys. Lett.* **65**, 97–99 (1994).

27. Unrau, W. *et al.* Electrically driven single photon source based on a site-controlled quantum dot with self-aligned current injection. *Appl. Phys. Lett.* **101**, 211119 (2012).

28. Kaganskiy, A. *et al.* Enhancing the photon-extraction efficiency of site-controlled quantum dots by deterministically fabricated microlenses. *Opt. Commun.* **413**, 162–166 (2018).

29. Unsleber, S. *et al.* Highly indistinguishable on-demand resonance fluorescence photons from a deterministic quantum dot micropillar device with 74% extraction efficiency. *Opt Express* **24**, 8539 (2016).

30. Somaschi, N. *et al.* Near-optimal single-photon sources in the solid state. *Nat Photonics* **10**, 340–345 (2016).

31. Jagsch, S. T. *et al.* A quantum optical study of thresholdless lasing features in high-β nitride nanobeam cavities. *Nat. Commun.* **9**, 564 (2018).

32. Shih, C.-W. *et al.* Low-threshold lasing of optically pumped micropillar lasers with $Al_{0.2}Ga_{0.8}As/Al_{0.9}Ga_{0.1}As$ distributed Bragg reflectors. *Appl. Phys. Lett.* **122**, 151111 (2023).

33. Liu, J. *et al.* Single self-assembled InAs/GaA quantum dots in photonic nanostructures: the role of nanofabrication. *Phys. Rev. Appl.* **9**, 064019 (2018).

34. Mai, L. *et al.* Integrated vertical microcavity using a nano-scale deformation for strong lateral confinement. *Appl. Phys. Lett.* **103**, 243305 (2013).

35. Ding, F., Stöferle, T., Mai, L., Knoll, A. & Mahrt, R. F. Vertical microcavities with high Q and strong lateral mode confinement. *Phys. Rev. B* **87**, 161116 (2013).

36. Engel, L. *et al.* Purcell enhanced single-photon emission from a quantum dot coupled to a truncated Gaussian microcavity. *Appl. Phys. Lett.* **122**, 043503 (2023).

37. Gaur, K. *et al.* High-β lasing in photonic-defect semiconductor-dielectric hybrid microresonators with embedded InGaAs quantum dots. Preprint at arXiv:2309.10936 (2023).

38. Gies, C. & Stephan Reitzenstein. Quantum-Dot Micropillar Lasers. *Semicond. Sci. Technol.* **34**, 073001 (2019).

39. Heermeier, N. *et al.* Spin-Lasing in Bimodal Quantum Dot Micropillar Cavities. *Laser Photon. Rev.* **16**, 2100585 (2022).

40. Birner, S. *et al.* nextnano: General Purpose 3-D Simulations. *IEEE Trans. Electron Devices* **54**, 2137–2142 (2007).

41. Mittelstädt, A., Schliwa, A. & Klenovský, P. Modeling electronic and optical properties of III–V quantum dots—selected recent developments. *Light Sci. Appl.* **11**, 17 (2022).

42. Yuan, X. *et al.* GaAs quantum dots under quasiuniaxial stress: Experiment and theory. *Phys. Rev. B* **107**, 235412 (2023).

43. Limame, I. *et al.* Epitaxial growth and characterization of multi-layer site-controlled InGaAs quantum dots based on the buried stressor method. Preprint at arXiv:2311.05777 (2023).

44. Burger, S., Zschiedrich, L., Pomplun, J. & Schmidt, F. JCMsuite: An Adaptive FEM Solver for Precise Simulations in Nano-Optics. in *Integrated Photonics and Nanophotonics Research and Applications* ITuE4 (OSA, 2008).

45. Panzarini, G. & Andreani, L. C. Quantum theory of exciton polaritons in cylindrical semiconductor microcavities. *Phys. Rev. B* **60**, 16799–16806 (1999).

46. Björk, G. & Yamamoto, Y. Analysis of Semiconductor Microcavity Lasers Using Rate Equations. *IEEE J. Quantum Electron.* **27**, 2386–2396 (1991).

47. Bennett, B. R., Soref, R. A. & Alamo, J. A. Del. Carrier-Induced Change in Refractive GaAs, and InGaAsP. *IEEE J. Quantum Electron.* **26**, 113–122 (1990).

48. Huang, H. C., Yee, S. & Soma, M. The carrier effects on the change of refractive index for n-type GaAs at $\lambda = 1.06$, 1.3, and 1.55 µm. *J. Appl. Phys.* **67**, 1497–1503 (1990).

49. LAMMPS molecular dynamics simulator. Available at: https://www.lammps.org/#gsc.tab=0. (Accessed: 7th November 2023).

50. Andreoli, L. *et al.* Optical pumping of quantum dot micropillar lasers. *Opt. Express* **29**, 9084 (2021).

# Supplementary Information

# Self-aligned photonic defect microcavities with site-controlled quantum dots

*C.-W. Shih,* [1] *I. Limame,* [1] *C. C. Palekar,* [1] *A. Koulas-Simos,* [1]

*A. Kaganskiy,* [1] *P. Klenovský,*[2,3,b)] *S. Reitzenstein*[1,a)]

[1]*Institut für Festkörperphysik, Technische Universität Berlin, 10623 Berlin, Germany*

[2]*Department of Condensed Matter Physics, Faculty of Science, Masaryk University, Kotlářská 267/2, 61137 Brno, Czech Republic*

[3]*Czech Metrology Institute, Okružní 31, 63800 Brno, Czech Republic*

Correspondence:

a) S. Reitzenstein, E-Mail: stephan.reitzenstein@physik.tu-berlin.de
b) P. Klenovský, E-Mail: klenovsky@physics.muni.cz

## 1. AFM profiles on the layer of SCQDs

Atomic force microscopy (AFM) scans are performed on SCQDs of a reference sample without the growth of top DBR. A bump around 8 nm is observed on most of the structures despite the surface morphology can change depending on the strain and the size of the oxide aperture as depicted in Fig. S1, originating potentially from Indium migration[1]. This bump acts as a monolithically integrated photonic defect after the growth of the top DBR and provides lateral mode confinement for vertical emitting photonic devices[2].

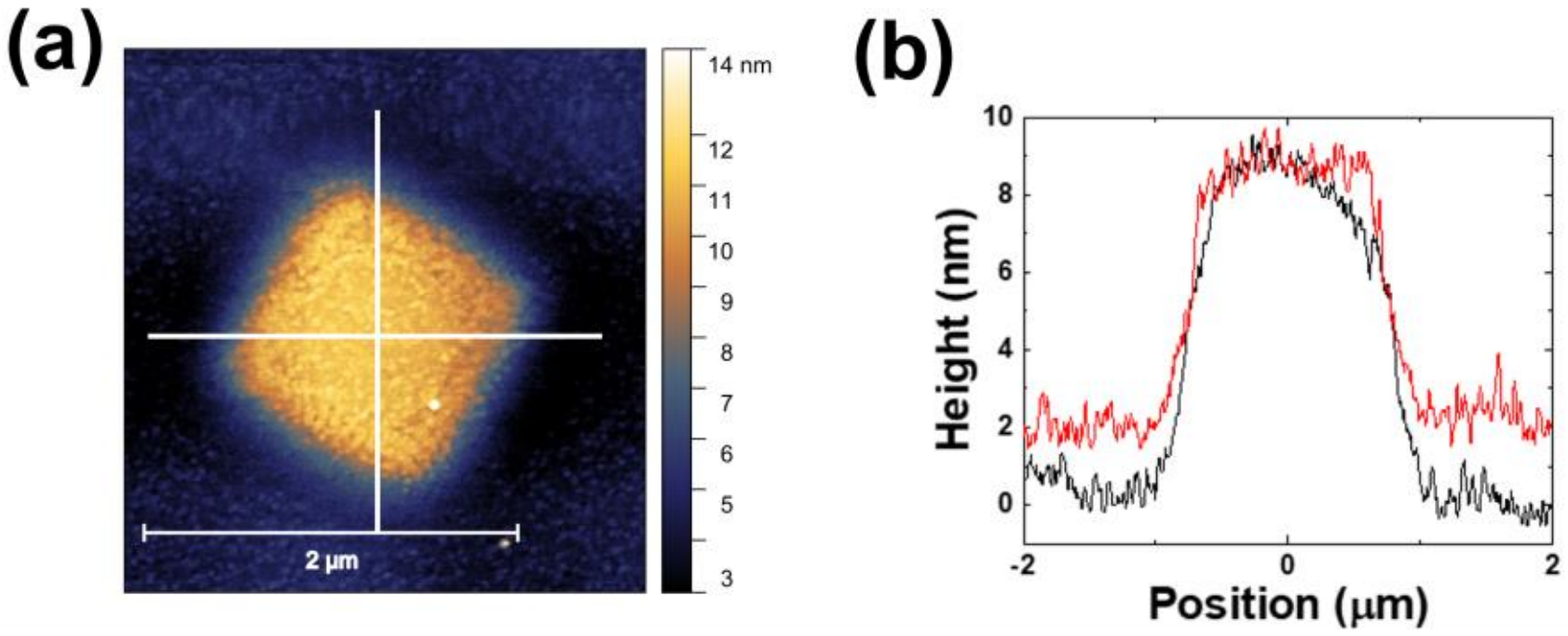

Fig. S1: (a) AFM profile of a structure with an oxide aperture diameter of 1.2 µm. (b) Two orthogonal line profiles of the AFM image indicated as the white lines in (a).

## 2. Optical simulation on and off the oxide area

In the main text, we describe that the two ensembles of photonic modes, including the aperture cavity modes and the mesa cavity modes, are observed in different spectral ranges. The aperture cavities are located directly above the center of the mesa, on the non-oxidized area. As a result, they contain a thicker optical cavity as proven by the AFM profiles and a non-oxidized AlAs layer with higher refractive index. That leads to an experimental redshift of the aperture cavity modes compared to the mesa cavity modes, the mode area of which consists of a large portion of the oxidized part.

To investigate those effects, we performed a 1-D optical simulation with transfer matrix method to verify the mode energy shift caused by the layer change on and off the oxidized-area. Fig. S2 shows the reflectance of the structure as a function of the wavelength, where the dips indicate the emission wavelength. A redshift of approximately 6.9 nm (9.9 meV) is observed in the simulations, which aligns with the experimental observations, specifically in Fig. 5 (a)-(c) of the main manuscript.

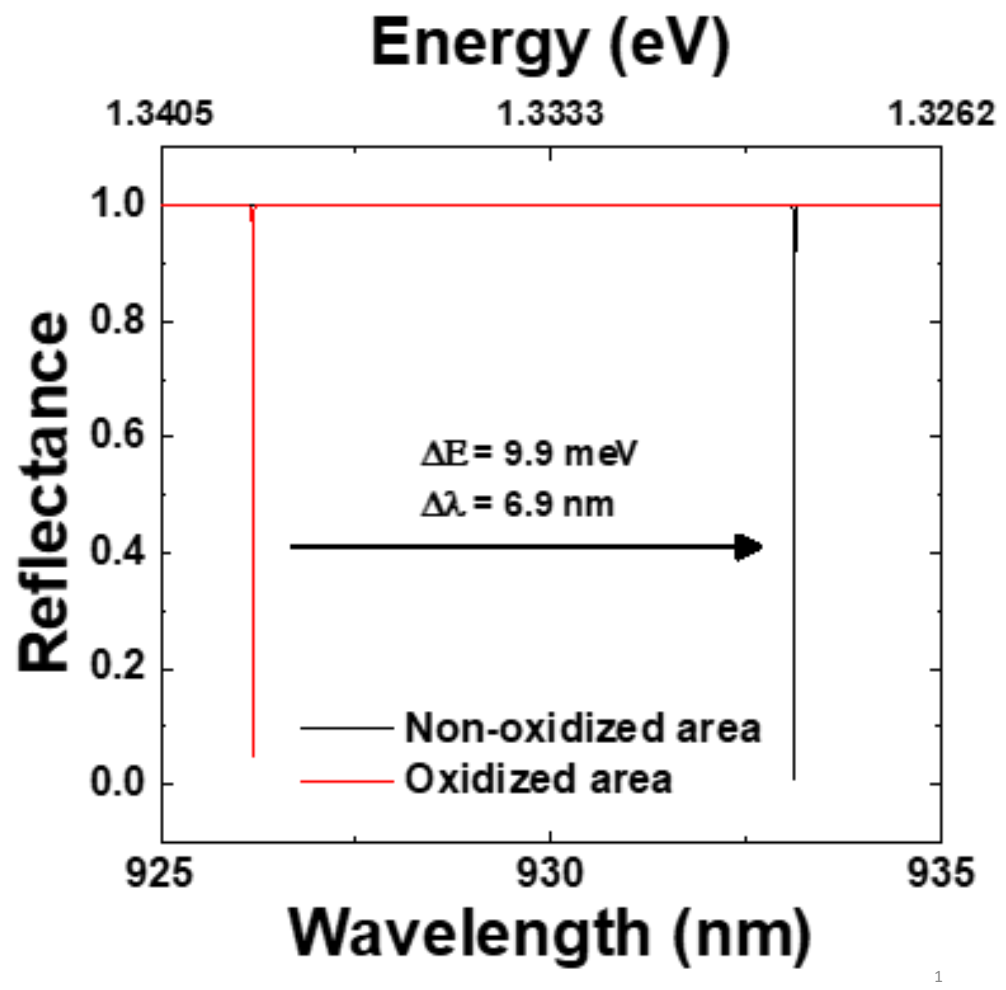

Fig. S2: Reflectance as a function of wavelength on and off the oxidized area. The reflectance dips indicate the emission wavelength of the fundamental cavity mode.

## 3. Optical measurements of mesa cavity modes

The optical measurements of a mesa cavity mode, where the contained SCM has a FWHM of 1.3 μm, are presented. Figure S3 (a) shows an exemplary emission spectrum of the mesa cavity mode with cavity Q-factor around 21000 at a pump power of 1 mW. With increasing excitation power, the mesa cavity mode enters the lasing regime, as we discuss below. Figure S3 (b) presents the polarization measurements above the lasing threshold, where the integrated intensity of the emission is plotted against the polarization angle, at a pump power of 40 mW. The mode exhibits a high degree of linearly. Figure S3 (c) and (d) present the excitation-power dependent measurements. Typical lasing characteristics including linewidth narrowing and a characteristic S-shaped IO-curve in double-logarithmic scale are observed. Similar to the aperture cavity modes, the effects of heating are absent, evidenced by the monotonous blueshift of the mode energy caused by plasma effects. By fitting using the laser rate equation, we extract a lasing threshold $P_{th}$ of (8.07 ± 0.43) mW and a β-factor of (2.31 ± 0.08) × $10^{-5}$. Noteworthy, the β-factor of mesa cavity modes is one order of magnitude smaller than the one of the aperture cavity modes presented in the main manuscript, which can be attributed to a larger mode volume and spatial mismatch of the SCQDs to the cavity mode antinode.

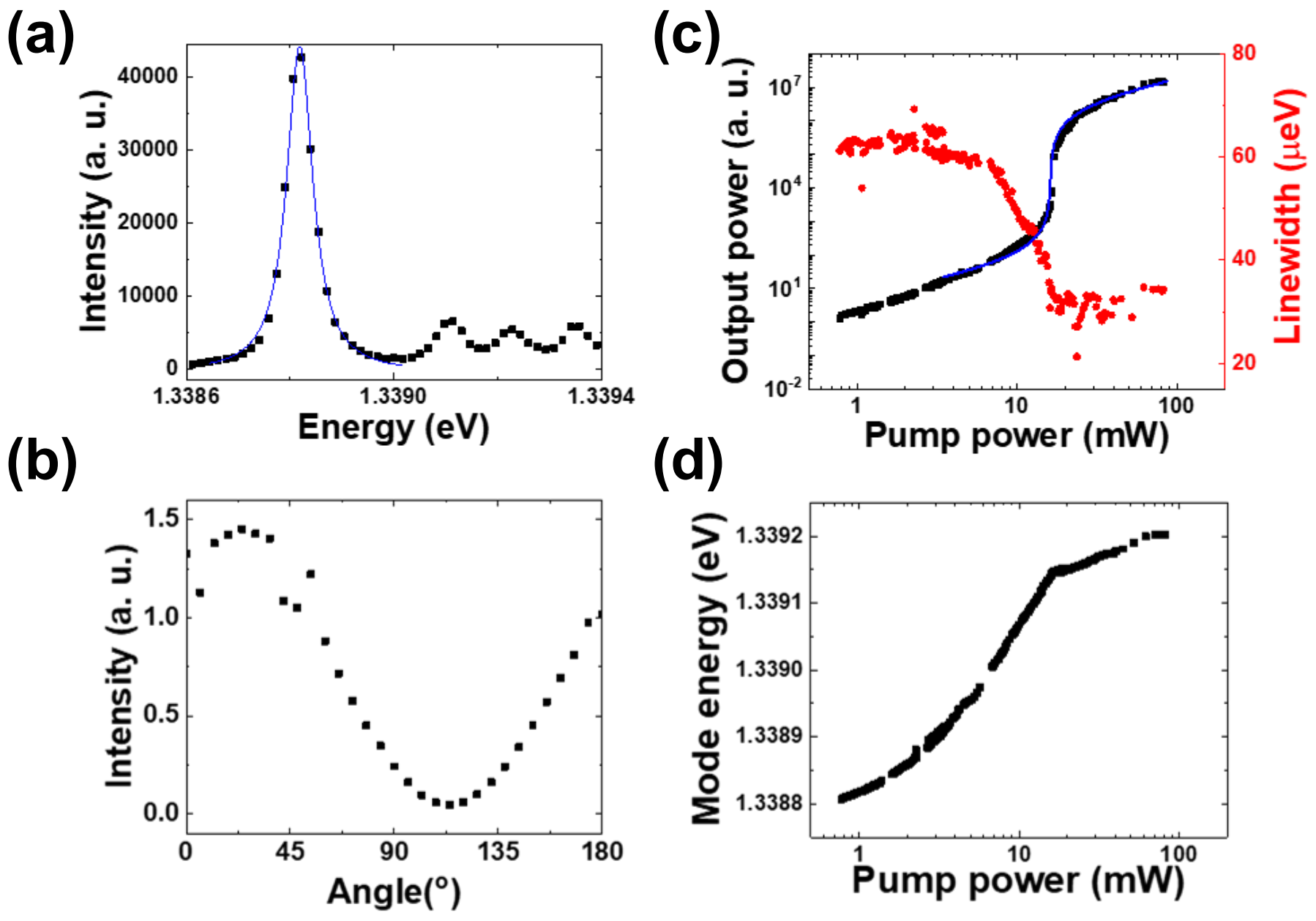

Fig. S3: Optical properties of a mesa cavity mode of an SCM with a FWHM of 1.3 µm. (a) Mode spectra of both polarization-orthogonal fundamental components at a pump power of 1 mW. (b) The integrated intensity as a function of the polarization angle components at a pump power of 40 mW above the lasing threshold. The mode is strongly linearly polarized according to the superior mode. (c) The IO curve and the associated mode linewidths. (d) The pump-power dependent mode energy.

## References

1. Limame, I. *et al.* Epitaxial growth and characterization of multi-layer site-controlled InGaAs quantum dots based on the buried stressor method. Preprint at arXiv:2311.05777 (2023).
2. Ding, F., Stöferle, T., Mai, L., Knoll, A. & Mahrt, R. F. Vertical microcavities with high Q and strong lateral mode confinement. *Phys. Rev. B* **87**, 161116 (2013).